%% file: manuscript.tex
\documentclass[twocolumn, a4paper,10pt]{article}
\usepackage[top=2.5cm, bottom=2.5cm, left=2.0cm, right=2.0cm,
columnsep=0.8cm]{geometry}
\usepackage{enumitem}
\usepackage[hidelinks]{hyperref}
\usepackage{boxedminipage}
\usepackage{nopageno}
\usepackage{graphicx}
\usepackage{natbib}
\usepackage[font=it]{caption}
\usepackage[usenames,dvipsnames]{xcolor}
\usepackage{listings}
\usepackage{caption}
\usepackage{subcaption}
\usepackage{graphicx} 
\usepackage{epsfig} 
\usepackage{amsmath} 
\usepackage{amssymb}  
\usepackage{multicol,multirow}
\usepackage{color}
\usepackage{xcolor}
\usepackage[export]{adjustbox}
\usepackage{graphicx} 
\usepackage{booktabs}
\usepackage{caption,subcaption}
\usepackage{natbib}
\setlist{itemsep=-0.1cm,topsep=0.1cm,labelsep=0.3cm}
\setenumerate{leftmargin=*}
\makeatletter
\renewcommand\title[1]{\gdef\@title{\fontsize{12pt}{2pt}\bfseries{#1}}}
\makeatletter
\renewcommand\section{\@startsection{section}{1}{\z@}{3pt}{3pt}{\normalfont\large\bfseries}}
\makeatletter
\renewcommand\subsection{\@startsection{subsection}{1}{\z@}{\z@}{\z@}{\normalfont\normalsize\bfseries}}
\makeatletter
\renewcommand\subsection{\@startsection{subsection}{1}{\z@}{\z@}{0.1pt}{\normalfont\normalsize\bfseries}}

\usepackage{scalerel}

\usepackage{fancyhdr}

\title{%
Benchmarking Model Predictive Control Algorithms  \\																								
\vspace{4pt}
in Building Optimization Testing Framework (BOPTEST)} 																																
\author{Saman Mostafavi$^{1}$, Chihyeon Song$^{2}$, Aayushman Sharma$^{3}$, Raman Goyal$^{1}$, Alejandro E Brito$^{1}$\\
$^{1}$ Palo Alto Research Center Inc (PARC), Palo Alto, CA, USA										
\\ 										$^{2}$ Korea Advanced Institute of Science and Technology, Daejeon, South Korea				
\\ 										$^{3}$ Texas A\&M University, College Station, TX, USA				
\\ 			 			  	
\\ 															
\phantom{Line 9}} 																																									
\date{\vspace{-0.5cm}}	
\begin{document}

\maketitle

\section*{Abstract}	
\addtocounter{section}{1}
We present a data-driven modeling and control framework for physics-based building emulators. Our approach consists of: (a) Offline training of differentiable surrogate models that accelerate model evaluations, provide cost-effective gradients, and maintain good predictive accuracy for the receding horizon in Model Predictive Control (MPC), and (b) Formulating and solving nonlinear building HVAC MPC problems. We extensively evaluate the modeling and control performance using multiple surrogate models and optimization frameworks across various test cases available in the Building Optimization Testing Framework (BOPTEST). Our framework is compatible with other modeling techniques and can be customized with different control formulations, making it adaptable and future-proof for test cases currently under development for BOPTEST. This modularity provides a path towards prototyping predictive controllers in large buildings, ensuring scalability and robustness in real-world applications.

\section*{Highlights}
\begin{itemize}
\item Development of a data-driven modeling and control framework for physics-based building emulators.
\item Offline training of differentiable surrogate models enhances model evaluations and provides cost-effective gradients for MPC.
\item Extensive evaluation of modeling and control performance across various BOPTEST test cases.
\item Compatibility with other modeling techniques and customizable control formulations ensures adaptability and future-proofing.
\item Practical implications for prototyping predictive controllers in large buildings, emphasizing scalability and robustness.
\end{itemize}

\input{tex/Introduction.tex}

\input{tex/SurrogateModel.tex}
\input{tex/ControlFormulation.tex}
\input{tex/Results.tex}
\input{tex/Conclusion.tex}

\bibliographystyle{bs2023}
\bibliography{references}

\end{document}

%% file: tex/Introduction.tex
\section*{Introduction}
According to recent estimates by \cite{doe/eia}, residential and commercial buildings account for nearly 40\% of energy usage in the United States. A significant amount of this energy consumption can be eliminated by improving the building's HVAC control system, for example, using predictive control methods as shown in \cite{drgovna2020all}. Among these methods, model predictive control (MPC) is a particularly powerful approach for handling constraints for state and control inputs in nonlinear multivariable control systems. While the gains are evident, the challenge is to show that MPC can be implemented at scale in a cost-friendly manner~\citep{o2022modelling}. It is well understood that the main obstacle to this is the modeling cost, and according to one study~\citep{atam2016control}, this can be as much as 70\% of the total effort of setting up an MPC-based building controller, mainly due to the effort and expertise required to create realistically calibrated models. Recently, the Building Optimization Testing Framework (BOPTEST) \citep{blum2021building} was developed to facilitate simulation-based benchmarking of building HVAC control algorithms. The emulator uses calibrated Modelica models to emulate building physical dynamics based on first principles. Models also output Key Performance Indices (KPI) that represent occupant satisfaction, energy cost and consumption, and carbon footprint. What makes this platform even more impressive is the fact that it is set up to replicate a real building control system with all its control limitations, e.g., there are realistic low-level feedback control laws, box constraints on control inputs, weather, occupancy profiles, economizer schedules, etc.

\textbf{Motivation:} While the value of BOPTEST, and other physics-based emulators in creating a unified testing platform for control, is unquestionable, there are several intrinsic obstacles to the implementation of predictive control and its adoption by a broader audience\footnote{It is worth mentioning that we consider these challenges to be almost identical for a real building HVAC control system and, therefore, addressing and solving them is a first step to deploying such control algorithms in the field.}: (1) In BOPTEST, and most other physic-based emulators, the numerical solvers for scaled-up models will not be computationally efficient to run in iterative optimization loops. (2) Solving the optimization problem requires gradient calculations which, derived through perturbations, only compounds the computational intensity. (3) Furthermore, some optimal control methods, such as iterative Linear Quadratic Regulator method (iLQR)~\citep{ilqg1}, derive optimal solutions by exploring trajectories that might be infeasible for the emulator to evaluate (due to control input and state constraints), which can lead to crashing the iterative algorithm prematurely.

While acknowledging the significant progress in deep neural networks-based reinforcement learning (RL) approaches, these methods are still highly data-intensive. The training time for such algorithms is typically very large, and high variance and reproducibility issues affect their performance \cite{henderson2018deep}. At the moment, RL algorithms remain intractable for adjustable and reproducible implementations at scale. On the other hand, most of the building MPC work~\citep{sturzenegger2015model,Mostafavi2022ifac,oei2020bilinear,mostafavi2019model} considers either simple low-fidelity RC-based models, bilinear models with low accuracy, Machine Learning (ML) approaches that cannot be directly used for fast MPC implementation, or directly uses Modelica-based models with hand-tuned cost functions for nonlinear optimization of energy consumption. Such modeling and control approaches require a lot of customization for high-fidelity models with complex, hybrid, and constrained systems that use external inputs and therefore, are not suited to a robust control framework.

\textbf{CONTRIBUTIONS}
This paper presents a modeling and control framework for building HVAC systems using differentiable models compatible with optimization-based nonlinear control methods. We propose a two-fold approach: (1) off-line identification of a differentiable surrogate model for the nonlinear mapping $x_{t+1}=f(x_{t},u_{t},d_{t})$, where $x$, $u$, and $d$ represent the state, control inputs, and external disturbances, respectively, and (2) employing automatic differentiation (AD)~\citep{paszke2017automatic} for solving nonlinear model predictive control (NMPC) with box constraints. We demonstrate the identification of suitable Neural Networks (NNs) and investigate various lags for states, controls, and disturbances. We present MPC formulations using AD, maintaining comfort constraints while minimizing KPIs for energy consumption. Our framework is customizable, adaptable to various control approaches, and provides warm-starting for MPC problems. We compare modeling and solving NMPC approaches and discuss best practices based on control criteria. We demonstrate NMPC control for the BOPTEST five-zone model, showing the scalability of our framework for data-driven NMPC control in building HVAC systems. To the best of our knowledge, the NMPC control of the BOPTEST five-zone model is the first of its kind.

%% file: tex/SurrogateModel.tex
\section*{Surrogate Modeling for Building Emulator}
\label{sec:model}
Our objective is to replace computationally expensive nonlinear numerical simulations with alternative, fast representations for model-based control. When using NNs for MPC, we propose incorporating the following criteria in the surrogate modeling process:

\begin{itemize}
\item{ \textbf{Computing cost:} Low computing cost for fast iterative evaluations.}
\item{ \textbf{Predictive accuracy:} High prediction accuracy for MPC's horizon.}
\item{ \textbf{Differentiability:} Fast and accurate gradient information for successive linearization, nonlinear solvers, etc., for various MPC formulations.}
\end{itemize}

We utilize the Pytorch~\citep{paszke2019pytorch} modeling library to achieve these goals. In this study, we consider Linear, MLP, and Long short-term memory (LSTM) models. MLP offers fast forward computation and strong expressivity to approximate complex functions \cite{hornik1989multilayer}. In contrast, since BOPTEST is a Partially Observable MDP (POMDP), it requires lag information from states, actions, and time-varying disturbances for model fitting. LSTM can address this by effectively handling nonlinear mappings with autoregressive features~\cite{siami2018comparison}. Although simple, the linear model provides the fastest model evaluations and plug-and-play compatibility for quick QP solvers.

\subsection{Linear}
The surrogate model takes states $x$, control inputs $u$, time-varying disturbances $d$, and their lags of past time-steps as input. The output is the future state prediction ${x_{t+1}}$:
\begin{equation}
    \begin{aligned}
   x_{t+1}=f(x_{t-M_{x}:t},u_{t-M_{u}:t},d_{t-M_{d}:t})
    \end{aligned}
\end{equation}

where $M_{x},M_{u},M_{d}$ are state, input and disturbance lags, respectively. Since the choices of lags are application dependent, we discuss this further in the result section. Here, $f$ is linearized as follows:
\begin{equation}
    \begin{aligned}
    x_{t+1} = \sum_{k=0}^{M_x}{A_k x_{t-k}} + \sum_{k=0}^{M_u}{B_k u_{t-k}} + \sum_{k=0}^{M_d}{C_k d_{t-k}}
    \end{aligned}
    \label{eq:nonlinear_map}
\end{equation}
where $A_k=\nabla_x f \in \mathbb{R}^{N_x \times N_x}, B_k=\nabla_u f\in \mathbb{R}^{N_x \times N_u}$ and $C_k=\nabla_d f\in \mathbb{R}^{N_x \times N_d}$ are learnable parameter matrices for state, control input and disturbance, respectively.

\subsection{MLP}
The linearized model given by Equation~\ref{eq:nonlinear_map} also applies here. The forward computation in MLP is written as the following:
\begin{equation}
    \begin{aligned}
    h_0 &= [x_{t-M_{x}}, u_{t-M_{u}}, d_{t-M_{d}}] \\
    h_{k+1} &= \tanh(W_k h_k + b_k), && k=\{0,..., K-1\} \\
    x_{t+1} = o_{t+1} &= W_K h_K + b_K
    \end{aligned}
\end{equation}
where $h_k \in \mathbb{R}^l$ is a hidden unit of the layer $k$, $W_k$ and $b_k$ are weight parameters of the layer $k$.

\subsection{LSTM}

The forward computation of LSTM is written as the following:

\begin{equation}
\begin{aligned}
    h_t, c_t &= MLP_{\text{enc}}(x_{t-M_x:t}, u_{t-M_u:t-1}, d_{t-M_u:t}) \\
    i_t &= \sigma(W_{ii}u_t + b_{ii} + W_{hi}h_{t-1} + b_{hi}) \\
    f_t &= \sigma(W_{if}u_t + b_{if} + Ww_{hf}h_{t-1} + b_{hf}) \\
    g_t &= tanh(W_{ig}u_t + b{ig} + W_{hg}h_{t} + b_{hg}) \\
    o_{t+1} &= \sigma(W_{io}u_t + b_{io} + W_{ho}h_{t} + b_{ho}) \\
    c_{t+1} &= f_t \odot c_{t} + i_t \odot g_t \\
    h_{t+1} &= o_t \odot tanh(c_{t+1}) \\
    x_{t+1} &= MLP_{\text{dec}}(h_{t+1})
\end{aligned}
\end{equation}
where $h_t$ is the hidden state, $c_t$ is the cell state, $i_t, f_t, g_t$ and $o_t$ are the input, forget, cell, and output gates, respectively. $\sigma(\cdot)$ is the sigmoid function, $\odot$ is the Hadamard product, and $MLP_{\text{enc}}$ and $MLP_{\text{dec}}$ are a MLP encoder and decoder, respectively.

%% file: tex/ControlFormulation.tex
\section*{Control Problem Formulation}
\label{sec:control}
Consider the discrete-time nonlinear dynamical system:
\begin{equation}
    x_{t+1}=f(x_t, u_t,d_t),
\end{equation}
where $x_t\in \mathbb{R}^{n_x}$ and $u_t\in \mathbb{R}^{n_u}$ correspond to the state and control vectors at time $t$ and $d_t\in \mathbb{R}^{n_d}$ is the set of contextual variables/external inputs. The optimal control problem is to find the optimal control policy that minimizes the cumulative cost:
\begin{align}
    & \min_{u_t} \sum_{t = 0}^{T} c_t(x_{t}, u_{t},d_{t}) \label{eq:Prb1} \\
    &\text{Subject to : }\  x_{t+1}=f(x_t, u_t,d_t), \label{eq:Prb2}\\ 
    &\text{Subject to : }\ u^l_t \leq u_{t} \leq u^u_t, \label{eq:Prb4}
\end{align}
for given $x_0$, and where $c_t(\cdot)$ is the instantaneous cost function given as:
\begin{equation}
\label{eqn:power}
\begin{aligned}
    c_t(\cdot) = P_c + P_h + L_k + \gamma P_x,
\end{aligned}
\end{equation}
where $P_c~\text{and}~P_h$ are total cooling and heating cost, $L_k =  \| \Tilde{u}_{t+1}-\Tilde{u}_{t}\|_R^2$ is a regularizer term, which penalizes large changes in the control inputs to avoid undesirable oscillations, and $P_x=\max(x_t^l - x_t, 0)+ \max(x_t - x_t^u, 0)$ enforces the occupant comfort constraints implemented with ReLU function with a penalty coefficient $\gamma$. The problem also considers input box constraints with lower and upper bound given as $[u^l_t,u^u_t]$.



\subsection{Gradient Descent Method}
The gradient descent method is one of the widely-used algorithms to optimize a differentiable objective function. At each iteration, the gradient of the objective function is computed and the decision variables are updated in direction of the computed gradient. Gradient descent algorithms have a precedent across domains such as training neural networks~\cite{schmidhuber2015deep} and solving optimal control problems~\citep{lin2014control}. In this paper, we use Adam~\citep{kingma2014adam}, which has shown promising results in deep learning applications. For input constraint (\ref{eq:Prb4}), we use projected gradient descent, a common method in solving constrained optimization: after each gradient update, we project the control vector $u_t$ into a feasible region $[u_t^l, u_t^u]$. Since the feasible region is a box constraint, the projected control vector is easily computed by using a clamp function after each update of the algorithm.

\subsection{Sequential Quadratic Programming}
There have been numerous tools and methods developed to solve specific nonlinear optimization problems with particular structures of cost functions, equality, and inequality constraint functions. However, Sequential Quadratic Programming (SQP) remains one of the most efficient approaches to solving any general constrained-nonlinear optimization problem. For the SQP approach, we utilize the optimization subroutine originally proposed by Dieter Kraft \cite{kraft1988software} and as implemented in SciPy~\cite{virtanen2020scipy} to solve the control optimization problem described in Eqns.~(\ref{eq:Prb1}-\ref{eq:Prb4}). The algorithm is a quasi-Newton method (using BFGS) applied to a Lagrange function consisting of a loss function and equality and inequality constraints. In our implementation, we provide the function evaluations, which are calculated using Equation~\ref{eqn:power}, and it's Jacobian using automatic differentiation. Instead of clamping, we pass bounds for control inputs directly to the solver.

%% file: tex/Results.tex
\section*{Results}
We demonstrate the effectiveness of our control framework for controlling building models in BOPTEST~\citep{blum2021building}, a software for simulation-based benchmarking of building HVAC control algorithms. This section details two test cases showcasing the results of deriving different surrogate models and discusses the subsequent control results for the control algorithms described in Section~\ref{sec:control}.

\subsection{Model Description}
BOPTEST emulators use Modelica~\citep{wetter2014modelica} to represent realistic physical dynamics. Embedded in these models are baseline control algorithms that can be overwritten using supervisory and local-loop control signals. BOPTEST uses a containerized run-time environment (RTE) which enables rapid, repeatable deployment of models. Using this feature, we stand up several instances of models on servers and query these models to speed-up data generation at scale for surrogate modeling. We also test controls on the same containers, representing \textit{digital-twins} of real buildings. We consider the following case studies:

\subsubsection{BESTEST Case 900 model}

This test case is a single room with floor dimensions of 6m x 8m and a floor-to-ceiling height of 2.7m. The building is assumed to be occupied by two people from 8 am to 6 pm each day. Heating and cooling are provided to the office using an idealized four-pipe fan coil unit (FCU). The FCU contains a fan, cooling coil, and heating coil. The fan draws room air into the HVAC unit and supplies the conditioned air back to the room. No outside air is mixed during this process. The fan has a variable speed drive serving the fan motor. The cooling coil is served by chilled water produced by a chiller and the heating coil is served by hot water produced by a gas boiler. Two different PI controllers for heating and cooling modulate the supply air temperature and fan speed to provide cooling and heating load to the room. For our supervisory MPC controller, we manipulate supply air temperature and fan speed as control inputs to minimize the combined cooling, heating, and fan power consumption while maintaining the occupant comfort bounds. Assuming the building to be in a climate close to Denver, CO, USA, the state and input box constraints are as follows:

\begin{align}
    & 21^o C \leq x^{T_{zone},occ} \leq 24^o C\\
    & 15^o C \leq x^{T_{zone},unocc} \leq 30^o C\\
    &  0.0\leq u^{fan} \leq 1.0 \\
    & 12^o C \leq u^{T_{supp}} \leq 40^o C 
\end{align}



 



\subsubsection{Multi-zone office (ASHRAE 2006 VAVReaheat)}
The test case represents the middle floor of an office building located in Chicago, IL, as described in the set of DOE Commercial Building Benchmarks for new construction~\citep{deru2011us} with weather data from TMY3 for Chicago O'Hare International Airport. The represented floor has five zones, with four perimeter zones and one core zone. The occupied time for the HVAC system is between 6 AM and 7 PM each day. The HVAC system is a multi-zone single-duct Variable Air Volume (VAV) system with pressure-independent terminal boxes with reheat. A schematic of the system can be found in \citep{blum2021building}. The cooling and heating coils are water-based, served by an air-cooled chiller and air-to-water heat pump respectively. A number of low-level, local-loop controllers are used to maintain the desired setpoints using the available actuators. The primary local-loop controllers are specified in \citep{blum2021building} as C1 to C3. C1 is responsible for maintaining the zone temperature setpoints as determined by the operating mode of the system and implements dual-maximum logic. C2 is responsible for maintaining the duct static pressure setpoint and implements a duct static pressure reset strategy. C3 is responsible for maintaining the supply air temperature setpoint as well as the minimum outside air flow rate as determined by the operating mode of the system. In this case, we assume the fan speed to be constant and our supervisory MPC controller manipulates the damper position and reheat control signal to control the airflow and zone supply air temperature respectively (at each zone). In addition, the central HVAC cooling and heating units are manipulated to control the central supply air temperature. The optimization objective is to minimize the overall cooling and heating loads while maintaining the occupant comfort bounds and central supply air temperature. The state and input box constraints are as follows\footnote{For more details on both case-studies, please refer to the figures and control mappings in \citep{blum2021building}}:

\begin{align}
    & 21^o C \leq x^{T_{zone_i},occ} \leq 24^o C\\
    & 15^o C \leq x^{T_{zone_i},unocc} \leq 30^o C\\
    &  0.0\leq u^{dam_i} \leq 1.0 \\
    &  0.0\leq u^{yReaHea_i} \leq 1.0\\
    \nonumber \\
    &  \forall i \in \{1,2,3,4,5\} \nonumber \\ 
    \nonumber\\
    &  5^o C\leq x^{T_{supp}} \leq 20^o C \\
    &  0.0\leq u^{yHea} \leq 1.0 \\
    &  0.0\leq u^{yCoo} \leq 1.0
\end{align}




    



    

\subsection{System Identification}
We consider the three choices of models as described in Section~\ref{sec:model} for the single zone and multi-zone case. We describe how we sufficiently excite the system to generate data and report the training and out-of-training performance of each model.\\
\textbf{Data generation}
For each time-step $t=0,...,T-1$, we sample a random control input $u_t$ from a uniform distribution of the feasible input space and pass the sampled control input to BOPTEST simulation to get the next observation and disturbance. We collect the data up to time-step $T$, and repeat this procedure $K$ times using different initial conditions. In the BESTEST case, we choose $K=120$, $T=500$, and use 100 distinct trajectories as training data, 10 for validation and 10 for test. In the multi-zone office case, we choose $K=600$, $T=1000$, and use 500 trajectories as the training dataset, and keep 50 for validation and 50 for test purposes. It is evident that test data, which all results are reported on, is the data that the model has never been trained on.\\
\textbf{Hyperparameters} The MLP framework consists of 4 layers with 256 nodes in each layer, and $\tanh(\cdot)$ activation layers in-between the MLP layers. For the LSTM model, we implement 2 layers with 256 nodes for $MLP_{\text{enc}}$ and $MLP_{\text{dec}}$ and choose the dimension of hidden and cell state as 256. Mean squared error (MSE) is used for computing training loss. For all surrogate models, we choose \textit{Adam} to optimize the parameters with learning rate=0.001, and epoch=1000.

\begin{table}
\caption{MSE~$(\times 10^{-5})$ for different model choices in BESTEST case}
\footnotesize
\centering
\setlength\tabcolsep{4.7pt} 
\begin{tabular}{l|ccc}
\toprule
\textbf{Model} & \textbf{Train MSE}  & \textbf{Val MSE} & \textbf{Test MSE} \\
\midrule
Linear & 699.5 & 566.8 & 780.3   \\
MLP & 8.846 & 12.70 & 17.56   \\
\textbf{LSTM} & \textbf{1.418} & \textbf{1.726} & \textbf{2.145}   \\
\bottomrule
\end{tabular}
\label{table:model_type}
\end{table}

\begin{table}
\caption{MSE~$(\times 10^{-5})$ for different MLP hyperparameter choices in multi-zone office case}
\footnotesize
\centering
\setlength\tabcolsep{4.7pt} 
\begin{tabular}{l|ccc}
\toprule
\textbf{$(M_x, M_u, M_d)$} & \textbf{Train MSE}  & \textbf{Val MSE} & \textbf{Test MSE} \\
\midrule
(1, 1, 1) & 511.6 & 623.9 & 618.6   \\
(1, 1, 5) & 476.0 & 623.8 & 624.3 \\
(1, 5, 1) & 20.46 & 21.74 & 24.35 \\
(5, 1, 1) & 82.43 & 98.92 & 103.8 \\
\textbf{(1, 5, 5)} & \textbf{14.71} & \textbf{17.76} & \textbf{18.47}   \\
(5, 1, 5) & 78.38 & 98.17 & 100.06 \\
(5, 5, 1) & 21.20 & 23.67 & 26.87 \\
\textcolor{red}{(5, 5, 5)} & \textcolor{red}{10.37} & \textcolor{red}{14.80} & \textcolor{red}{14.82} \\
\bottomrule
\end{tabular}
\label{Table:MLP}
\end{table}

\begin{figure}[b]
    \centering
    \includegraphics[width=1.0\linewidth]{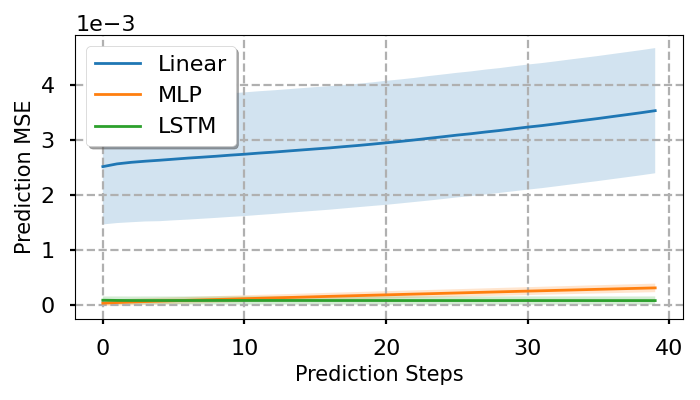}
    \caption{Test MSE for different choices of surrogate models in multi-zone test case. LSTM and MLP have comparable performance and outperform the Linear model.}
    \label{fig:multizone_prediction_mse}
\end{figure}

\textbf{Predictive performance}
Table~\ref{table:model_type} and Table~\ref{Table:MLP} show the results of test performance for single-zone and five-zone models respectively. Losses are calculated using average prediction error for 40 steps.
For multi-step ahead prediction, a \textit{for}-loop is implemented in the forward propagation of the ML models. The results for single-zone and multi-zone models demonstrate the superiority of LSTM in prediction accuracy, although, MLP performance is comparable in the five-zone case as depicted in Figure~\ref{fig:multizone_prediction_mse}. 

\begin{figure*}[ht!]
    \centering
    \includegraphics[width=1.0\linewidth]{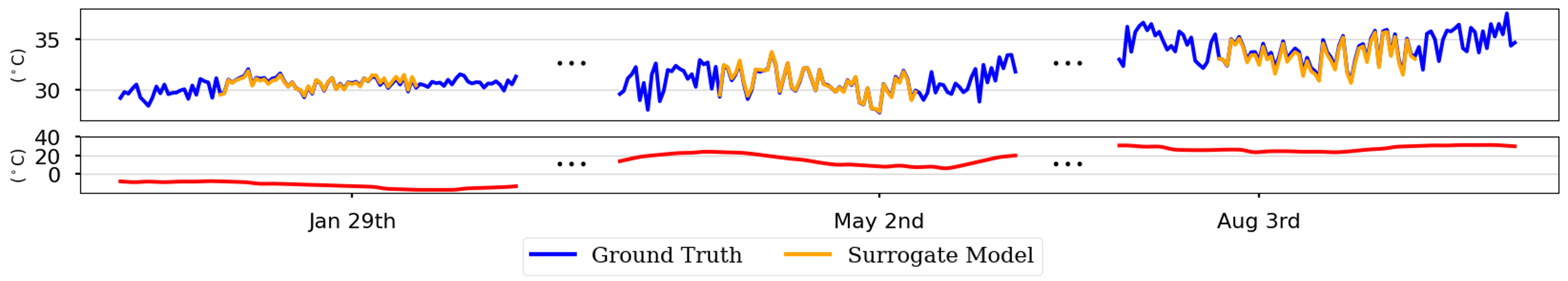}
    
    \caption{The set of figures show the results of out-of-training predictive performance for five zone model during three distinct weather events (January, May, and August) for core zone (top). The ambient temperature trajectories is depicted in red (bottom). The orange lines represent the 50-step ahead predictions (12.5 hours) starting from the left most point of the trajectory. The full MSEs are reported in Table~\ref{Table:MLP}.}
    \label{fig:multizone_prediction_plot}
\end{figure*}

In Table~\ref{Table:MLP}, we compare the performance of different MLP model choices with different lag values of the state, input, and time-varying disturbances. (5,5,5) is the best model among all choices but (1,5,5) model comes very close with fewer model inputs. This model depends on lags of weather data and control inputs, which we speculate is not unrelated to the lags associated with lower-level controllers in this system. We chose (1,5,5) as a more simple, equally accurate choice.
Figure~\ref{fig:multizone_prediction_plot} is a visual depiction of the predictive accuracy of the chosen MLP for surrogate modeling of the five-zone model during three distinct weather events (January, May, and August) for the core zone. Each orange trajectory is a 50-step ahead prediction (12.5 hours) starting from the leftmost point of the trajectory. These results appear to be conclusive for deploying the model in MPC.

\subsection{Control Results}
For all control algorithms, we deploy a receding-horizon controller, wherein a 10-step "look-ahead" trajectory is generated using the optimization algorithm, and only the first step of the optimization solution is passed to BOPTEST model to obtain new measurements. The new data point is then used as the initial condition for the next iteration of the control optimization. 
In addition, to speed up convergence, the previously optimized control trajectory is used as the initial trajectory for warm-starting the receding horizon replanning for the MPC problem.

The control results for single-zone and multi-zone cases are reported in Table~\ref{Table:result_1} and Table~\ref{Table:result_2}, respectively. In the single-zone case, LSTM model performs best for control. This is expected from the superior predictive accuracy of the model. It also has the best average computation time. As for the control algorithm, Gradient-based approach finds a better local minima for the problem. In the multi-zone case, LSTM performs poorly (unexpectedly) and MLP outperforms all models. Here, in contrast to the previous case, SLSQP finds a better local minima. Next, we discuss the significance of these results. 

\subsection{Discussion}
The modeling results indicate that it is possible to derive accurate ML models from the building emulators. It is worth mentioning that the bottleneck in this process is data generation which is not always trivial for hybrid systems with many if-else conditions, low-level control loops and system constraints, and finely-tuned numerical solvers. 

On the control side, we have conducted extensive tests using SLSQP and Gradient-based approaches from various initial conditions. In the one-zone case, the gradient-based approach with the LSTM model demonstrates the lowest power consumption while maintaining an acceptable discomfort level. However, in the multi-zone case, SLSQP with the MLP model achieves the lowest power consumption, even though the LSTM model exhibits better predictive performance. We hypothesize that the more complex architectures and increased number of parameters in LSTMs, as compared to MLPs, can make the Jacobian calculations more challenging. The LSTM's memory cell introduces additional nonlinearities and feedback loops, which can increase the risk of numerical instabilities, potentially affecting the quality of the Jacobian calculations. This, in turn, can diminish the control performance. This observation somewhat contradicts the expectation that better predictive performance always guarantees better control performance. Based on this experiment, we believe that a balance between model complexity and predictive performance should be considered for these types of MPC problems. Alternatively, improved control formulations might help address this issue. As there is limited precedent in the literature, we are conducting more tests to seek better and more definitive answers. It is worth noting that the framework is functioning as intended, enabling the formulation of new hypotheses based on experimentation.

\subsubsection{Computation Time}
By comparing the average computation time between several methods, we make the following interesting observations: First, both the gradient-based approach and SLSQP show comparable computation time, though the computation time of both solvers depends on their stopping criteria. For example, after running extensive tests, we decided that 100 iterations was a good stopping criteria for the gradient-based approach. We expect this hyperparameter tuning to be problem specific. Second, for the surrogate model, it is obvious to us that MLP should take longer than the Linear model to run. Surprisingly, the LSTM model, which has the most complex structure among the three candidates, shows the fastest computation time. We think that this computation time gap most likely comes from a difference in the implementation language. Each surrogate model has a $for$-loop to predict the multi-steps. Although all surrogate models are implemented in Pytorch, the linear and MLP model conduct their $for$-loops in python, while LSTM model uses C++.


\begin{table}
\caption{Average of total power$(kWh/m^2)$, thermal discomfort $(kh/\text{zone})$ and computation time $(sec)$ on BESTEST case}
\footnotesize
\centering
\setlength\tabcolsep{4.7pt} 
\begin{tabular}{ll|ccc}
\toprule
\textbf{Model} & \textbf{Solver} & \textbf{Power}  & \textbf{Discomfort} & \textbf{Time}\\
\midrule
Linear & GDM & 0.0189 & 1556 & 1.607 \\
Linear & SLSQP & 0.2551 & 1528 & 0.933 \\
MLP & GDM & 4.804 & 2.935 & 1.694 \\
MLP & SLSQP & 5.059 & 5.207 & 1.684 \\
\textbf{LSTM} & \textbf{GDM} & \textbf{4.818} & \textbf{2.081} & \textbf{0.620} \\
LSTM & SLSQP & 4.943 & 4.415 & 0.661 \\
\bottomrule
\end{tabular}
\label{Table:result_1}
\end{table}

\begin{table}
\caption{Average of total power$(kWh/m^2)$, thermal discomfort $(kh/\text{zone})$ and computation time $(sec)$ on multi-zone office case}
\footnotesize
\centering
\setlength\tabcolsep{4.7pt} 
\begin{tabular}{ll|ccc}
\toprule
\textbf{Model} & \textbf{Solver} & \textbf{Power}  & \textbf{Discomfort} & \textbf{Time}\\
\midrule
Linear & GDM & 2.807 & 10.44 & 1.504 \\
Linear & SLSQP & 2.487 & 11.40 & 1.600 \\
MLP & GDM & 3.458 & 4.054 & 1.782 \\
\textbf{MLP} & \textbf{SLSQP} & \textbf{2.778} & \textbf{3.154} & \textbf{2.144} \\
LSTM & GDM & 2.222 & 124.7 & 0.570 \\
LSTM & SLSQP & 2.880 & 35.48 & 0.818 \\
\bottomrule
\end{tabular}
\label{Table:result_2}
\end{table}

\begin{figure*}[t!]

\begin{multicols}{2}

\subfloat[]{\includegraphics[width=\linewidth]{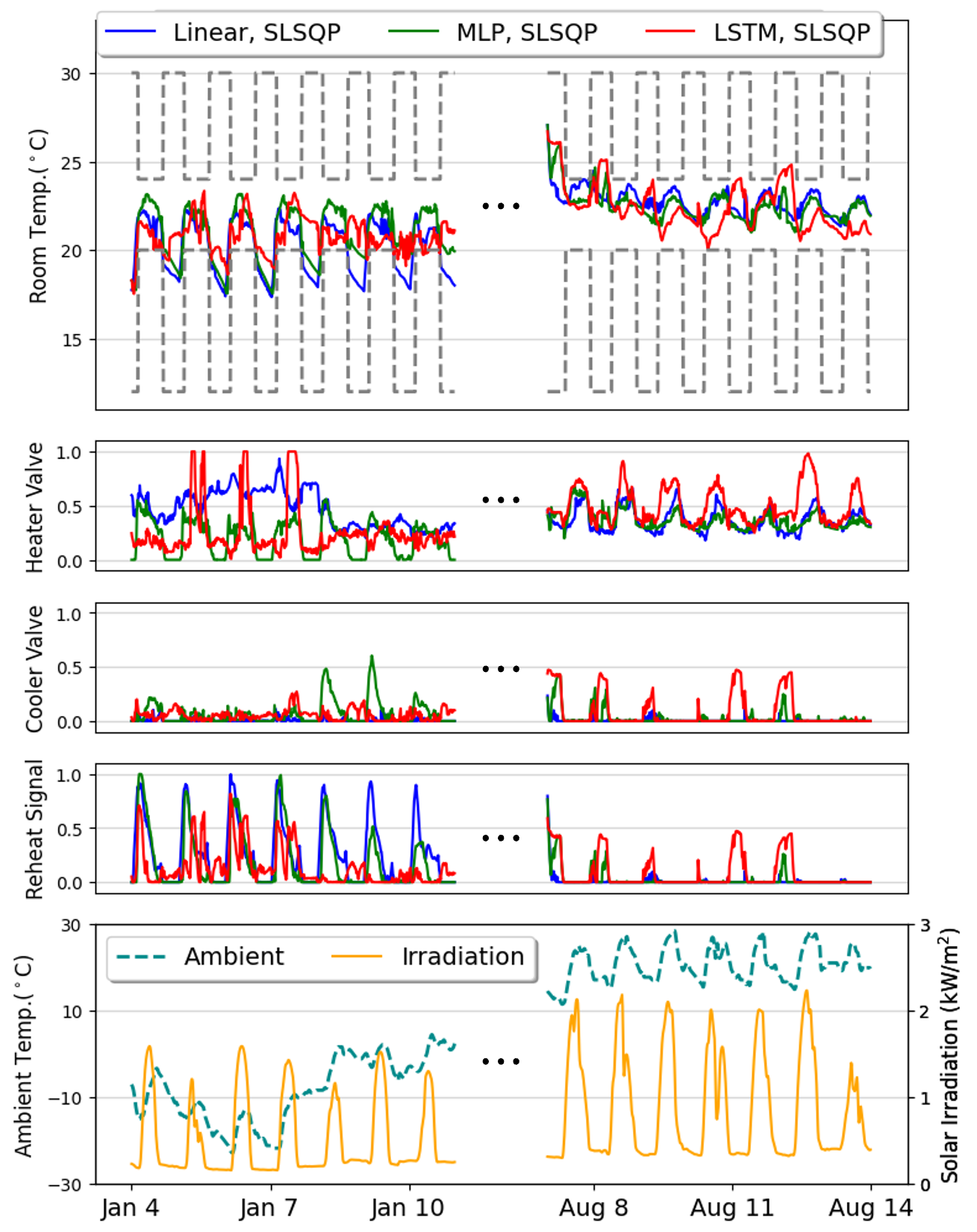}}

\subfloat[]{\includegraphics[width=\linewidth]{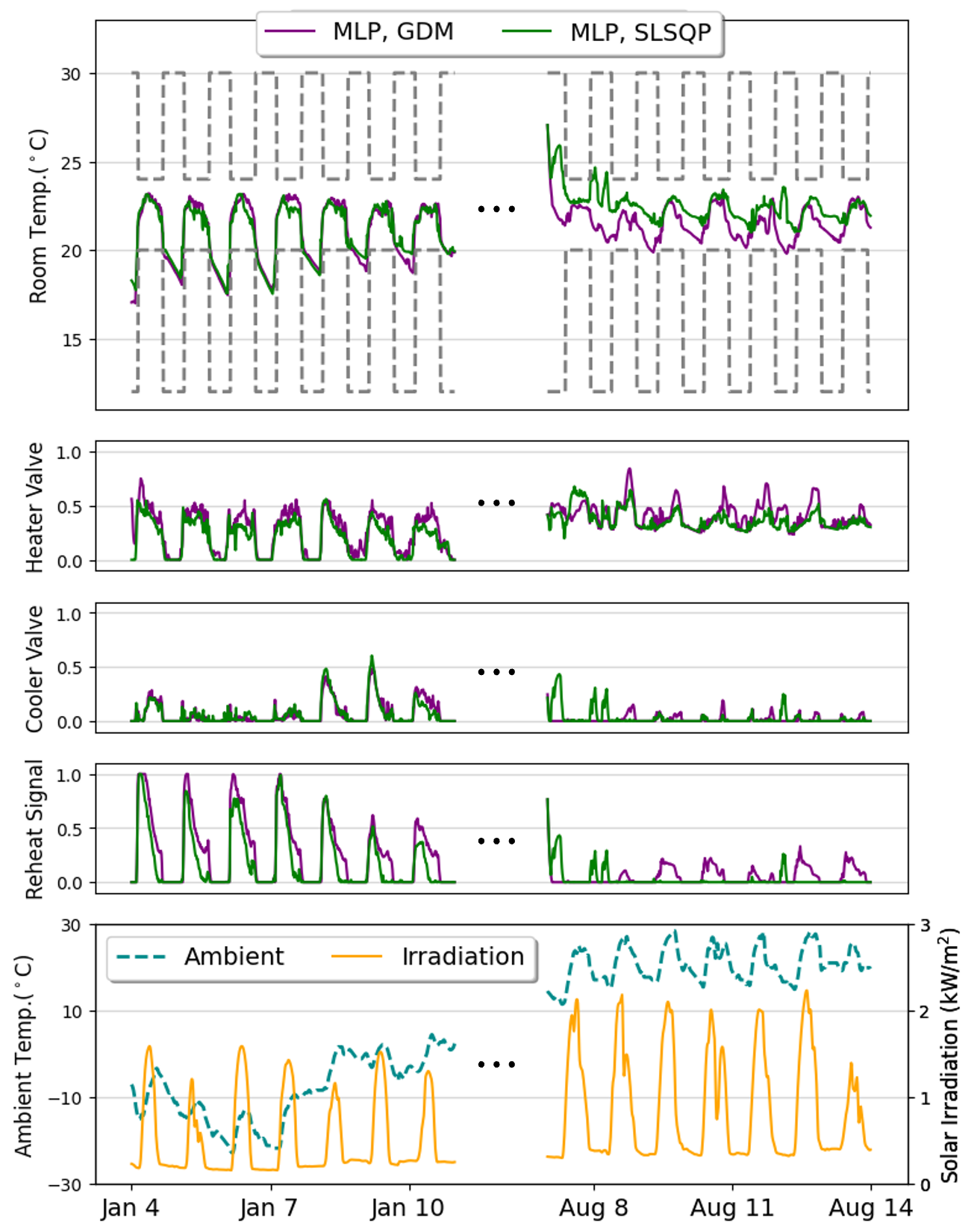}}

\end{multicols}
       \caption{Result comparison for different choices of models and control algorithms. The top figure represents the temperate. The bottom figure is the relevant weather data, and the middle figures are the corresponding control inputs. The results are divided into a cold (Jan) and hot (Aug) weather events. (a) Result for control of core-zone in the multi-zone test case using SLSQP with Linear, MLP, and LSTM models. Using MLP model, the control outperforms LSTM and Linear model-based implementation. (b) MLP-based control results with SLSQP solver slightly outperform the Gradient-based approach.}
    \label{fig:multizone_result}
\end{figure*}

%% file: tex/Conclusion.tex
\section*{Conclusion and Future Work}
We presented a modeling and control framework for controlling physics-based building emulators. We have shown that our approach is successful in reducing cooling and heating loads in the BOPTEST emulator while satisfying occupant comfort and adhering to control system constraints. The approach is modular, meaning that it will be compatible with various other choices of models and control algorithms. For example, while we did not succeed in training a good LSTM model for the five-zone case, we anticipate that the right hyperparameter tuning should address this issue and we are actively working on it. The same is true for control. For example, we tested the framework with an iLQR controller which failed to satisfy constraints. While we did not manage to get the results we expected, we anticipate that significantly better control results are possible with iLQR and we are currently fixing our implementation of the algorithm. This is especially important since iLQR has shown superior performance for nonlinear optimal control problems~\citep{ilqg2}. We are also exploring other fast first-order solvers with alternative control formulations. For example, we are considering OSQP~\citep{stellato2020osqp}, which will significantly speed up the optimization while producing high-quality solutions, or distributed ADMM~\citep{boyd2011distributed} for district-level problems. In addition, we are actively working with the developers of BOPTEST to control scaled-up models, including multiple coupled buildings, with the framework. 

The main bottleneck for scaling the current approach is the customized nature of the data generation process. In the current process, many trials and errors are required to find a feasible input space that does not break the emulator in forward simulations. Latest studies\citep{chakrabarty2022simulation} provide some promising insight into more robust sampling procedures. We are currently working on incorporating similar approaches into our process. 

Last but not least, while in this paper we focused on control as an application, we firmly believe that system design, fault diagnosis, and reliability are other applications that will benefit from the proposed modeling approach, and we are actively investigating problems in these domains.